\documentclass[preprint,authoryear,5p,times]{elsarticle}
\usepackage{amssymb}

\journal{ }

\begin{document}

\begin{frontmatter}

\title{Open star cluster: formation, parameters, membership and importance}

\author[label1]{Gireesh C. Joshi}
\address[label1]{P. P. S. V. M. I. College, Nanakmatta, Udham Singh Nagar, Uttarakhand-262311}

\begin{abstract}
We have been represented the collective information of estimation procedures of parameters of the open clusters and put them together for showing the importance of clusters to understand their role in stellar evolution phenomenon. Moreover, we have been discussed about analytic techniques to determine the structural and dynamical properties of galactic clusters. The members of clusters provide unique opportunity to determine their basic parameters such as, age, metallicity, distance, reddening etc. The membership probability of stars of clusters is assigned through the various approaches and each approach provides different number of probable members of the cluster. Here, we have been briefly discussed about various approaches to determine the stellar membership within clusters.
\end{abstract}

\begin{keyword}
Open star cluster---Spatial parameter----Dynamical--Mf--LF----Membership

\end{keyword}

\end{frontmatter}

\section{Introduction}
\label{intro}
The observational foundations of stellar astrophysics are studies of the Sun and stellar clusters \citep{cu13}. Open star clusters (OCs) or Galactic open clusters are gravitationally bound system of stars formed together (i.e. similar identified age) and their stars are at practically the same distance with similar chemical composition whereas stars in a single cluster are differing only in mass \citep{ha05}. Hence, they are dynamically associated system of stars which are formed from giant molecular clouds through bursts of star formation \citep{ma08}. The study of an individual star provides only limited, frequently inaccurate and uncertain information about it and possibly about the interstellar medium whereas, OSCs provide an ideal opportunity to simultaneously study a group of stars located in a relatively small space \citep{ze12}. Furthermore, it is believed that mostly stars in the Galaxy are formed in open clusters \citep{la03}, therefore, OCs are key probes for tracing the star-forming history (SFH) \citep{da12}. Such objects are excellent laboratories for studying the structure, kinematics and evolution of the Galactic disc \citep{ya11} with chemical composition \citep{fr95}, consquently prescribed as the main building blocks of the stellar populations in our Galaxy \citep{kh13}. These information of different locations of Galaxy are prescribed about Galactic evolution processes in the different time span \citep{pa10}. Overall, OCs are found to be true representatives of the Galactic disk population, systematic investigation of their nature, size, number of members, and age \citep{kh05}. Furthermore, these systems are assigned Solar orbits around the Milky Way centre \citep{zh15}.
 
Beside these, many questions remain unanswered, such as the location of inner spiral arms or determination of the co-rotation distance \citep{gl13}. These OCs are studied  to understand the space-age structure, formation and evolution of the Galactic disk \citep{ch07} and the stellar evolution as well \citep{ta11}. They are believed important test beds for star formation, stellar initial mass function (IMF) and stellar evolution theories \citep{ga08}. Mostly OCs are poorly studied or even unstudied up to now \citep{pi11}. Their study is of great interest and attractive work in several astrophysical aspects and verification of their evolution theories. The Galactic, radial and vertical, abundance gradient can be studied by OCs  \citep{ho00,ch03,ki05}. In addition, these objects are good tools to analyze the large-scale proprieties of the disk of our galaxy and testing tools of the theories of stellar and galactic evolution \citep{ta13}. Hence, their observational studies have become a traditional benchmark for testing our comprehension of several aspects of the formation and properties of the Galactic disk \citep{mo10,ca10}.
 These studies are possible due to fact that their stars emerge from the same molecular cloud. The OCs have also provided constraints on the understanding of the step-like abrupt decrease in metallicity at 1 kpc from the Sun \citep{le11}.

A substantial number of OCs is hidden behind the interstellar material in the Galactic plane \citep{fr07}. Researchers could be identified new clusters due to availability of data of huge infrared (IR) surveys and a low interstellar extinction in the IR wavelength range. This low value of extinction gives an opportunity to deeply dig into the spiral arm areas where most of the OCs are concentrated and more than a thousand OCs were discovered by analysing the Two Micron All Sky Survey (2MASS) \citep{kr06,fr07,ko08,gl10}. This survey has proven to be a powerful tool in the analysis of the structure and stellar content of OCs \citep{bi03}. This survey is also useful for investigating the star formation process and the stellar evolution \citep{sh05}.

The structural parameters of OCs hold important information about mass like basic parameters and surrounding galactic tidal field \citep{ho57}. The position of a cluster centre and apparent (angular) size are needed to identify and describing an OC on the sky \citep{pi07}. These common parameters are present in numerous catalogues of available for about 1700 OCs \citep{di02,kh13,di14}. On the basis of their ages, they have divided into three groups such as Young, intermediate and old-age OCs. The study of Young clusters provide information about current star formation processes and are they also becomes key objects for describing questions about Galactic structure, whereas observational studies of old and intermediate-age open clusters play an important role to study of theoretical evolution aspects of galaxy \citep{sh06}. The Young OCs are effectively used to understand the nature of interactions between young stars and interstellar medium due to fact that they are still embedded in the parent nebulous regions \citep{sc04}.

Their members are categorized in various groups such as main sequence (MS), Pre-main sequence (PMS), etc. As PMS stars become older, they contract towards the zero-age main sequence (ZAMS) and their core temperatures rise \citep{je13}.
The study of dense central region (the core) as well as the expanded and sparse region (the halo or corona) is necessary to understand cluster evolution \citep{ma09}. The core contains relatively bright and massive ($\geq$~3 $M_{\odot}$) stars along with low-mass stars \citep{br99}, whereas the corona contains a large number of faint and low-mass ($\leq$~1 $M_{\odot}$) stars. These regions has an important bearing on studies related to the mass function (MF) and the structure of OCs \citep{sh06}. A detailed structural analysis of coronae of OCs is needed to understand the effects of external environments, like the Galactic tidal field and impulsive encounters with interstellar clouds, etc., on dynamical evolution of OCs \citep{pa90}.

The radial density profile (RDP) gives the picture of stellar densities within various concentric radial zones which can be used to identified the sparse (poorly populated) and rich clusters. The former OCs do not survive longer than a few hundred Myr , whereas later ones may be survive longer \citep{pa86,ca05,bo12}. The fundamental physical parameters of OCs, e.g. distance, reddening, age, and metallicity are play an important role to study about the Galactic disk.  The depth of photometric analysis of OCs directly depends on the statistical completeness of the sample, especially in the faint-GC tail, and on the reliability of the derived parameters \citep{bo08}. The colour–colour (CC) diagrams of an OCs is valuable tool for studying interstellar extinction in the directions of clusters \citep{ya04}. The colour–magnitude diagrams (CMDs) of photometric studies of clusters allow us to estimate their fundamental parameters \citep{ya08}. The described diagrams of OCs usually show a well-defined long and broad main sequence (MS) \citep{du01}. Typically, the determination of basic parameters of OCs via isochrone fitting requires either that the metallicity is estimated, or that an priory value is adopted \citep{ol13}. Therefore the metallicity becomes a required parameter for the precise determinating of parameters of the OCs. Though, many times solar metallicity is often assumed due to the complexity of the required observations and requirement of very indirect methods for its determination \citep{ol13}. High resolution spectroscopy is often claimed to be the most accurate method for determining the metallicity value of clusters \citep{mag09}, though the value of metellacity of individual member of dense cluster could not be possible through present instruments due to their resolution powers. We also note that individual metallicity studies might well be able to achieve a high precision for a limited number of stars \citep{me09}.

The quantitative comparisons between theory and observations of OCs suffer either obscuring material due to their location in the Galactic plane or expected many field interlopers of given region \citep{su99b}. The small-scale structure, large-scale structure and observational errors produce distortion in redshift which leads to difficulties in determining the real cluster members \citep{ab11}. Membership selection is therefore of crucial importance in the study of open clusters. The more precision cluster parameters could be computed through the membership analysis. It is well known fact that the proper motions (PMs) of stars of an OC provide a unique opportunity to obtain membership information for these stars \citep{ya13}. These said proper motion values of the stars would be extracted from the online available catalogues \citep{ro10,za13}. In recent years, the detailed membership analysis of stars in the cluster field has become a subject of intense investigation, mainly in view to understand the cluster properties \citep{ca08,ya08,jo14,jo15a}. In nearby clusters (d $<$ 200 pc), the average large motions of cluster members from background stars makes easy identification of them \citep{khb13}.

Mass spectrum of these objects can be used to study the initial mass function (IMF) \citep{la14}. The shape of the initial mass function (IMF) is effective parameter to drawn a theory about the fragmentation of molecular clouds and therefore the formation and development of stellar systems \citep{sa00}. The statistical analysis of the spatial distribution of the probable cluster members and their mass function can be useful to drawn some information about their dynamical evolution \citep{an02,ka02}. Mass segregation and evaporation of low-mass members are the main effects of dynamical interactions \citep{fu01,ca06} and these interactions are becomes probe to understand the evolving phenomena of the clusters. In the process of evolution, a significant fraction of the cluster stars is gradually lost to the field. Dynamical evolution and mass segregation can have a significant effect on the radial structure as well as on the shape of the present-day mass function (MF) of open clusters. The MF of OC is usually obtained from the observed luminosity function (LF), which can be obtained from the observed CMDs \citep{sa03}. The universality of the IMF of OCs is still an open question due to elementary considerations about the IMF i.e. it depend on star-forming conditions \citep{la98}. Present estimations of the observed IMF do not constrain the nature of the IMF \citep{kr07}. Identifying systematic variations of the IMF with different star-forming conditions would allow us to study early cosmological events \citep{kr02}. IMF has similar properties in very different environments and should depend only on the process of molecular cloud fragmentation \citep{kr02,bo06}. The difficulties such as complex corrections for stellar birth rates, life times, etc associated with determining the mass function (MF) from field stars are automatically avoided due to similar age and metaliicities of the members of OCs \citep{sh08}.

The OCs are continuously harassed by external processes aged such as tidal stress and dynamical friction (from the Galactic bulge and disk and from giant molecular clouds), and by internal ones, such as mass loss by stellar evolution, mass segregation, and low-mass star evaporation \citep{la05,kh07,bo08}. The most OCs are disrupting on a timescale of a few hundred Myr due to Galactic gravitational tidal forces \citep{so10}, therefore most clusters seems to be relatively young. The other hand, the last stage of the evolutionary model of the clusters is believed to be a cluster remnant \citep{jo15b,pa11}.
\section{Star Cluster formation}
The analysis of observational data becomes indicators to constrain the model of stars formation within the clusters. In a cloud of hydrogen gas, laced with helium and a trace of other elements, something triggers to a physical phenomena collapse, which leads Stars formation processes of a range of different masses and begin their lives. The gravitational pull of
the parent galaxy will disrupt the clusters and these objects present a snapshot of those stars, which have formed more or less at the same time and at the same place. As a result, these clusters are studied by the astronomers with the great interest.\\ 
The theory of stellar structure and evolution within cluster is based on the mass dependent evolution of stellar luminosities and temperatures with age. The dating of young, embedded clusters is considerably more difficult due to the majority of PMS stars in the embedded clusters and inaccuracy in the measurement of their luminosities and temperatures. The embedded-cluster phase of evolution lasts somewhere between 3 and 5 Myr \citep{la10}. Other hand, the MS of older clusters terminates at fainter and cooler levels. The changes of the stars during the end of their lives may be constrains by differences observed among stars away from the MS. Star formation processes is occurred in two modes as spontaneous star formation and triggered star formation.
\subsection{Spontaneous star formation}
The tendency of the interstellar matter of galactic scale is to be condense under gravity into star-forming clouds, which is counteracted by galactic tidal forces. The star formation is take place for that dense cloud, for which, its self gravity to overcome these tidal forces. In the spontaneous star formation (SSF), the huge molecular cloud convert into a protostar by falling inward towards a centre i.e. collapse process of said cloud. This prescribed collapse occurs due to the instability of huge cloud. In the earlier SSF theories, it was believed that it is the result of the gradual cooling of the warm interstellar gas, bringing the thermal Jeans mass to a value lower than the mass of the cloud. A cold cloud may be born by turbulence or magnetic tension and pressure or rotation. Moreover, collapse of cold cloud occurs due to loses its turbulent energy via any dissipation mechanisms or loosing of magnetic field diffusion or disappearance of cloud’s angular momentum due to magnetic tension. The matter accretion of cloud may be occurred due to the some combination of these three energies leads instability and mass change-without any loss of energy \citep{el98}.
\subsection{Triggered star formation}
A significant low mass stars are formed in molecular clouds in the vicinity of massive luminous stars \citep{he05}. The interaction of $UV$ radiation from $OB$ stars with their surrounding molecular gas environments is necessary to study about star fromation processes and hydrodynamics of complex dynamic structures \citep{le94}. This formation is take place due to the presence high pressure of gases. These mechanism is divided into three types, which are given as: (i) Strong, (ii) Moderate and (iii) Weak triggering. A diffuse clouds is strike  with a gravitational/radiation shock which leads stars formation through the compressed diffuse clouds. This phenomena is known to be {\bf Strong} triggering. Such type star formation is not possible through the other alternating phenomena. The {\bf Moderate} triggering is increased the star formation rate in that molecular cloud region where stars already forms through the spontaneous star formation. Thus, the shock increased the compression rate of gas in the {\bf Moderate} triggering which leads faster rate of star formation compare to the spontaneous star formation. The {\bf Weakly} triggering occurs due to the bulk motion of the gas, which leads varying star formation region with the bulk motion of the gas. Furthermore, star formation rate remains constant in this triggering.\\
Triggering is also divided into three groups according to the size of star formation region/rate \citep{el98}. The {\ small scale triggering} involves the direct squeezing of pre-existing cloud or globules by high pressure (nearly surrounds the whole cloud). An {\bf intermediate scale triggering} is the compression of a cloud from one side. Furthermore, a dense ridge of gas eventually collapses or recollects into denser cores in this triggering leads star formation. The {\bf large scale triggering} is an accumulation of gas into an expanding shell or ring which partially surrounds the pressure source, a star or supernova. The all type of triggering may be result of a gravitational instability due to the density enhancement of a shock moving through a molecular cloud. Recurrence of triggered star formation results in sequential star formation \citep{fu00}. These triggering processes are occurred due to the various external agents. The brief discussion of these agents are prescribed as below,
\subsubsection{{\bf Supernovae:} ``Death of massive stars trigger new generation of stars"}
The supernovae are those massive stars, in which explosions are occurred due to the stellar evolution. These explosions are giving rise to enormous amount of radiation to their surroundings leads to initiate star formation in nearby clouds by their supernovae-driven
shock waves. Numerical studies \citep{va98,fu00} show that shock-waves with velocities in the range of $15-45~km ~s^{-1}$ corresponding to supernova explosions at a distance between 10 pc and 100 pc from the molecular cloud can induce such a collapse.
\subsubsection{{\bf Massive O/B-type stars :} ``Expansion of H II region"}
Massive stars emit very large amount of ultraviolet (UV) radiation. The surrounding gas region of these objects is ionized and heated due to the interaction of UV radiation. Furthermore, such type regions are defined as the {\bf H II regions}. Since, these regions are contained higher temperature (~$ 10,000$ K) and higher pressure compare to its surrounding neutral hydrogen regions, therefore H II region expands with a typical velocity of the order of ~$11~km ~s^{-1}$. Furthermore, its expansion rates decreases with time \citep{dy97}. H II region may undergo with various events during the expansion \citep{de05}. These events are given by various theories/models as follow: (i) Radiation driven implosion i.e. RDI process \citep{che07}, (ii) collect and collapse model \citep{el77}, (iii) instability of ionization front, (iv) interaction between an H II region and a nearby filamentary molecular cloud \citep{fu00} and (v) supersonic turbulent molecular medium \citep{el95}.
\subsubsection{Spiral arm}
Spiral arms of the Galaxy are prominent regions of the ongoing star formation and fulfilled by the presence of young stars, H II regions, dust and giant molecular clouds \citep{el83}. The said fulfilled matter within arms is the indication of the stars formation processes. The fixed angular speed of rotation of enhanced density regions of spiral arms is defined by the pattern speed and said regions are considered to be waves of compression, i.e. density waves \citep{li64}. The speed of density waves is found to be high for the inner part of the Galaxy with reference to the arms, whereas low for the outer part of said arms of the Galaxy. A large cloud may sudden compressed due to its interaction with the spiral arm density leads increment of local mass density with the collapse of cloud \citep{li64}. The large scale distribution of molecular clouds of spiral arms can be generated through the interaction between spiral shocks and cold atomic gas \citep{do06}.
\section{Spatial parameters of the cluster}
Since, the cluster parameters are highly depends on the observed photometric stellar magnitudes of a particular portion of the sky or field of view (FOV) of the cluster region. The cluster existent may be probably understood through the crowding behaviour of the stars within the FOV of the observed sky region. Though, clusters are not having circular shape but  it is assumed to be circular for determining the cluster radius in a simple way. A general discussion about determining procedure of the various spatial parameters of the cluster as given below,
\subsection{Way of determination of the center of the cluster}
For determining the cluster radius, it is mandatory to determine the cluster center. It is assumed that the center of cluster is a that point in which the stellar density becomes maximum. The star count method \citep{sa98} is a best method for determining the center of any cluster. Though it does not guaranteed fact that the point, having maximum most stellar number/density, shows exact cluster characteristics \citep{jo15b}. In the case of above prescribed fact, we would left maximum stellar point and the second maximum point would considered to be center of the cluster. As a result, the finding procedure of the center becomes iterate procedure until the exact cluster characteristics does not appear. Furthermore, the radial stellar densities are used to define the stellar crowding effect within the cluster instead of stellar number. The stellar densities are used to overcome the condition of blank/dark regions of the cluster. Such blank/dark regions appear in the small portion of the cluster, however, such effect would be disappears in the case of large scale structure. This fact is well matched with the homogeneity property of the large scale structure of the Universe.\\
The above prescribed radial densities are estimated within various radial zones. If, a total of $N_{i}$ stars are found within any radial zone (having area $A_{i}$) then the radial density of that zone estimated by the relation as: ${\rho_{i}=\frac{N_{i}}{A_{i}}}$. The width of radial zone may be kept in the manner that it does not affected cluster characteristics. The smaller width of the zone of any cluster produces highly varied densities values due to the presence of undefined dark/blank regions within the cluster. The high width of radial zone reduce the precision of estimation value of the cluster radius. Furthermore, the said width of the radial zone changes according to the stellar density of the studied cluster. The smooth cluster characteristics may be obtain for low radial width of the dense cluster compare than the width of the sparse cluster. Thus the resultant profile of radial stellar densities are played an important role to determine the cluster extant. 
\subsection{Statistical models to determination of the cluster radius}
The radial densities profile (RDP) of any cluster contains two axis. The x-axis represents the radial distance, whereas y-axis represents  the radial densities. Since, it is believed that the field stars are symmetrically distributed in the nearby sky of the cluster, therefore, the cluster region is contained the stellar enhancement with respect to the field region. On this background, it is necessary to consider the field stellar density ($\rho_{bg}$) within the cluster. This said density must be reduced from the computed stellar densities of the various radial density. There are various statistical formulas to compute the radius of the cluster through these modal radial densities i.e. $\rho_{r}-\rho_{bg}$ ($\rho_{i}$ is the observed stellar density of the $i^{th}$ radial zone). The most applied formula for estimating the center density ($\rho_{0}$) and core radius ($r_{c}$) of the cluster is King-Empirical formula \citep{ki66,ka92} as given below, $$\rho_{r}-\rho_{bg}=\frac{\rho_{0}}{1+\left(\frac{r}{r_c}\right)^2}.$$ 
The said core radius of the cluster is defined as the radius where the density, $\rho_{r}$, is half of the central density, $\rho_{0}$ \citep{pa08}. Other hand, the radius of the cluster is defined as the distance from the centre to the point where the stellar density of the cluster merged with the field region. Since, it is noticeable fact that the best fitted curve of the  King-Empirical model and model background density i.e. $\rho_{bg}$, are never intercept to each other. Since, this fact leads to an infinite extent of the cluster and produce a contradiction to compute the cluster extant, therefore the field stellar density is defined as the sum of 1-$\sigma$ error with the $\rho_{bg}$. Thus, the radius of the cluster is the distance from the center of the cluster to interception point between the best fitted curve and field stellar density. Though, the 1-$\sigma$ error is the statistical prediction for estimating the cluster radius and the resultant radius would be changed with significance level of the estimation such as 1-$\sigma_{bg}$ or 3-$\sigma_{bg}$. On this background, the radius of the cluster is also estimated through the estimation error in background stellar density as follow,
$$r_{cluster}=r_c \sqrt{\frac{\rho_0}{\sigma_{bg}}-1}.$$
Since, the above prescribed radius depends on the estimation error in the $\rho_{bg}$, therefore, it may be varied according to said estimation error. This error depends on the smoothness of the radial density profile, which further depends on the width of the assumed radial zone. The width of radial zone is selected by researcher on the behalf of his own sense. Thus, resultant radius may not representing the true radius of the clusters. It is believed that the gravitational attraction among stars of the cluster is vanished beyond the cluster periphery, however it does not satisfied the reasons of sudden disappearing of attraction. Though, the cluster members are shown the evaporating nature from the cluster region, so the disappearance of the said weakly gravitational bound may occurred before the end of extent of the cluster. Thus, the gravitational bounding among stars of the cluster might not hold together in the adjacent stellar enhancement region of field region. As a result, it is needed to define a radial distance to describe the possible gravitational bounded region of the cluster and such region is defined by the limiting radius. This limiting radius is a radial distance, after which no stars are gravitationally bound to the cluster system \citep{jo15b}. This said limiting radius is computed by the following relation \citep{bu11},
$$r_{limiting}=r_c \sqrt{\frac{\rho_0}{3 \sigma_{bg}}-1}.$$
Though, the estimation error of the field stellar density is also influenced the value of said radius, however computed value might to be reliable in the sense that the said value with uncertainty lies within the stellar enhancement region. On the basis of limiting radius, the maximum borderline background stellar density i.e. $\rho_{bg_{max}}$ is estimated by the following relation,
 $$\rho_{bg_{max}}=\rho_{bg}+3\sigma_{bg}.$$
This limiting radius is also used to compute the concentration parameter ($c$) following as \citep{pe75},
$$c=log \left(\frac{r_{limiting}}{r_{c}}\right).$$
This relation is directly indication of positive value of $c$ for that stellar system, in which the limiting radius is greater than the core radius. Furthermore, this parameter may be characterized the density profile of each cluster in such a way that it is independent of its diameter and richness.
\subsubsection{Tidal radius}
Besides above prescribed radii, we have found tidal radius of the cluster, which is computed through the total cluster mass ($M_{cl}$) as follow \citep{bi87},
$$R_{t}=D_{G}\left(\frac{M_{cl}}{3M_G}\right)^{1/3}.$$
Here we have adopted a Galactocentric distance $D_{G}\thickapprox8~kpc$ and a Milky Way-like galaxy with an enclosed mass of $M_G=5.8 {\times} 10^{11} M_{\odot}$ \citep{zh15}. Thus, the tidal radius is estimated using the relation given as $R_{t} = 1.46 (M_{cl})^{1/3}$ \citep{je01}.
\subsubsection{Limitations of RDPs}
The various limitation of RDPs is described as below \citep{nil02},\\
(i) The RDPs is dependent on the limiting magnitudes  of the detected stars. It seems that the cluster radius may be increased with the limiting magnitude of detected stars.\\
(ii) In some cases, the weak contrast between cluster and field region in outer region creates inaccuracy for defining the angular size of the cluster.\\
(iii) The exact cluster boundaries has not been defined due to the irregular shape of the cluster.\\
(iv) The identification procedure of the centre seems to be difficult for the sparse and poor cluster due to presence of clump in RDP.\\
(v) The cluster members becomes diminishes with the increasing radial distance due to inability of detection them from the statistically fluctuation of the field stars.
\subsubsection{Surface number density}
The surface number density $\sigma(r,m)$ is defined by the number of stars per magnitude interval in unit area. Furthermore, it is a function of magnitude and radius from the cluster center \citep{su96}. The SNDPs are very effective to estimate the cluster extent and the nature of stellar density within various magnitude bins \citep{jo16}. These profiles are highly effected due to the presence of data incompleteness and stellar crowding effect. Furthermore, the characteristic curves of this profile are overlap to each other at the periphery of the cluster. As a result, these profiles are least used to determine the cluster extent. Beside this, these profiles of clusters are the signature of stellar distribution of various masses leads the revolutionary information about the stellar evolution processes within the clusters.
\subsection{Ellipticity of the cluster}
Generally, the cluster radius has estimated by considering the spherical shape of the cluster but cluster shape is non-spherical. This non-spherical shape is occurred due to the galactic tidal field \citep{wi85}, differential rotation of the Galaxy, encounters with giant molecular clouds \citep{gi06}. Thus, the semi-major ({\it A}) and semi-minor ({\it B}) axes of the apparent distribution of cluster members gives ellipticity $e$ following as,
$$e=1-\frac{B}{A}$$
The semi-axes are considered to be the principal axes of an ellipse using characteristic equations for the second-order moments of tangential coordinates of the 1$\sigma$ cluster members \citep{pi08}. However, this results may be influenced by the membership procedure of the cluster.
\subsection{Various methods of the reddening determination}
The reddening of the cluster is a colour-excess but every colour-excess has not representing this value. This prescribed reddening value has computed through the observed stellar magnitudes in $B$ and $V$ photometric bands. It is represented by the $E(B-V)$. The solution values of the best fitted isochrone of $(B-V)~vs~V$ CMD has provided the its value following as,
$$(B-V)_{ob}=E(B-V)+(B-V)_{in}.$$
where $(B-V)_{ob}$ and $(B-V)_{in}$ are the observed and intrinsic colour-excess values for detected stars. The above said reddening value may be obtained by any one method from the following described approaches. OScs are very frequently affect by the large colour excesses and this affection may be $10 \%$ of the extinction across the whole cluster region \citep{ta13}.
\subsubsection{Reddening through $(B-V)$ vs $(V-I)$ TCD}
The solution of linear fit to the main-sequence (MS) in the $(B-I)$ versus $(B- V)$ plane of the cluster may obtained through the following relation \citep{ca02},
$$(B-I)=Q+2.25{\times}(B-V)$$
Those MS stars are selected for the linear fit in above relation, which have $V-$magnitude less-than or equal to 18 mag. The resulting value of $Q$ from the above relation would be used to determine the reddening by following Munari \&  Carraro \citep{mu96} relation,
$$E(B-V)=\frac{Q-0.014}{0.159}.$$
This above prescribed relation is applicable for the $R_{V}=3.1.$ Furthermore, this relation has used in the certain colour-range as $-0.23 \leq (B-V)_{ob} \leq +1.30$.  
\subsubsection{Reddening through (U-B) vs (B-V) TCD}
The reddening, $E(B-V)$, in the cluster region is estimated using the $(U-B)$ vs $(B-V)$ TCD. To determine the reddening, we have to take only those stars which have spectral type earlier than $A_{0}$ as later type stars may be more effected by the metallicity and background contamination \citep{ho03}. In this approach, we would be added some shifted values on the colour-excesses [$(U-B)_{in}$ and $(B-V)_{in}$] of the zero age-main sequence (ZAMS, \citep{sc82}), these shifting values are chosen in such a way that they must be satisfied the following relation \citep{de78},
$$\frac{E(U-B)}{E(B-V)}=0.72.$$ 
The error in reddening is calculated using the following relation \citep{ph94},
$${\sigma^2_{{(B-V)}_{sys}}}={\sigma^2_{{(U-B)}_{o}}}+{\sigma^2_{{(B-V)}_{o}}}+{\sigma^2_{centre}}.$$
where $\sigma_{vector}$ has considered to be 0.01.
\subsubsection{The reddening through the 2MASS colours}
The $(J-H)$ vs $(H-K)$ diagram of each cluster provides the value of the ratio, $\frac{E(J-H)}{E(H-K)}$ through best fitted isochrone and these values are satisfied the relation, $\frac{E(J-H)}{E(H-K)}=\frac{5}{3},$ \citep{me70} in case of normal reddening law. The reddening, $E(B-V)$ is estimated using the following relations \citep{fi03},
$$E(J-H)=0.309E(B-V)~~~and~~~E(J-K_{s})=0.48E(B-V)$$
where $E(B-V)$ is the reddening.
\subsubsection{Interstellar extinction in near-infrared}
The study of interstellar extinction in the NIR is effective to examine the nature of interstellar matter/dust. A $(V-K)$ vs $(J-K)$ diagram is also used to determine the interstellar extinction in the near-IR (NIR) range. We have been added colour-values in such a way that they are approximately satisfied the following given relation, $E(J-K)/E(V-K)=0.173.$ Finally, reddening value can be computed through the relation $R_{V}=1.1E(V-K)/E(B-V)$ \citep{wh80}. These relations are also used to examine the nature of reddening law towards to the clusters region. If, the resultant reddening value through this approach is well matched with the reddening value through $(U-B)$ vs $(B-V)$ TCD, then extinction law is said to be normal. Generally, normal reddening law is applicable when dust and intermediate stellar gases are absent in the line of sight of the cluster \citep{sn78}. In addition, the extinction law is found to be normal for solar neighbourhood, the Galactic Centre or in moderately young open clusters and associations \citep{ri85,gu89,li11}. If, the values are not matched to each other, then it becomes indication of the presence of stellar dust/gases. On this background, dust evolution of the star-forming regions becomes one of the more interesting issues in the astronomy due to direct dependency of grin size with the reddening law \citep{li14}. 
\subsection{Absorption relations and nature of reddening law}
The apparent stellar magnitudes are found to be different for different photometric bands. Since, the stars are radiating the stimulated radiation in every bands, therefore, the absolute magnitudes can be computed through the these different apparent magnitudes. It is believed that absolute magnitude of a star is a fixed value in any band, which leads to co-relate the observed colours value with prescribed absolute magnitude. Thus, the absorption relations are defined by those transformation relations/equations, which are used for converting the absorbed colour values into absolute magnitudes. The brief notes of these relations are given below,
\subsubsection{Various relations}
In the view of $R_V=3.1$, the correction in reddening values have been carrying out through absorption relation \citep{du02} as below,
$$A_J/A_V=0.276,~A_H/A_V=0.176,~A_{K_S}/A_V=0.118,$$ $$and~A_J=2.76{\times}E(J-H).$$
A total selective absorption of the stars is estimated through the reddening as follow \citep{ol75,wa87},
$$R=A_{V}/E(B-V)=3.06{\pm}0.25 (B-V)_{0} + 0.05 E(B-V).$$
Similarly, the $(B-V)$ and $(V-I)$ colour excesses may be related as follow \citep{wa85},
$$E(V-I)=1.25E(B-V)[+0.06(B-V)_{0}+0.014E(B-V)].$$
For simplicity, the mean intrinsic colour i.e. $(B-V)_{0}$, of the clusters is taken to be 0.0 mag. 
\subsubsection{Extinction Law}
As the light emitted from the star cluster, it passes through the interstellar dust and gas, hence the scattering and absorption of light takes place. Since the absorption and scattering of the blue light is more compared to the red light, the stars appear reddened. The reddening law is expected to be different in the presence of dust and gas and its nature is investigated by using $(V-\lambda)/(B-V)$ TCDs \citep{chi90}, where $\lambda$ is any broad band filter other than $B$ and $V$. The $(V-\lambda)/(B-V)$ TCDs have been widely used to separate the influence of the extinction produced by the diffuse interstellar material from that of the intra-cluster medium \citep{chi90,jo14}. Therefore, a best linear fit in the resultant $(V-\lambda)/(B-V)$ TCD diagram of each cluster gives the value of slope ($m_{cluster}$) for the corresponding TCD. Assuming the value of $R_{normal}$ for the diffuse interstellar material as 3.1, a total-to-selective extinction $R_{cluster}$ is determined as follow \citep{ne81},
$$R_{cluster}=\frac{m_{cluster}}{m_{normal}} \times R_{normal}.$$
The relation $\frac{\lambda - V}{B-V} = R_{cluster} (a_{\lambda} - 1) + b_{\lambda} $ \citet{ca89} is also used to determine the unknown values of colour-excess through above prescribed value of $R_{cluster}$.
\subsection{Distance and age}
Theoretical isochrones gives opportunities from the simple age-dating of star clusters, to more complex problems like the derivation of star formation histories of resolved galaxies \citep{mar08}. A classical isochrone of a fixed metallicity can be represented by a single line in the Hertzsprung-Russell diagram \citep{br11}. These prescribed isochrones are providing the distance and age of the cluster by their best fitted solution on the various CMDs. The broadness of stellar scattering is mainly caused by the presence of binaries and field stars in the cluster region. It is also noticed that the prescribed broadness of the fainter members are not reducing after the field star decontamination through the statistical cleaning procedures. There are three key issues in the determination of the distance and age of the clusters as follow, (i) the accurate metallicity of the cluster, (ii) the exact location of the turnoff point on the main sequence and (iii) the visual inspected solution of the best fitted isochrone.\\
The distance and age of the clusters are determining by keeping the fixed reddening and metallicity values. The exact value of the metallicity of the cluster are computing through the spectral analysis of its members spectra.
\subsection{Isodensity contour}
The morphology of the clusters can significantly influence by internal interaction of two-body relaxation, which are result of encounters among member stars and external tidal forces either due to the Galactic disc or giant molecular clouds \citep{pa08}. In the case of young clusters, the initial morphology should be governed by the initial conditions in the parent molecular cloud  \citep{ch04}. Isodensity contour of a cluster is defined by that region, for which, obtained stellar density is found to be fixed value. Since, the stellar density of a cluster is varied from center to its periphery, therefore, the isodensity contour are containing a region, in which, stellar density is found between limits of any given density bins. The contour of higher density seems to be embedded within the contour of lower density. It is also appeared that the contours of higher densities are spread out in the cluster and not connected to each other, whereas outer boundary of other contours are shown similarity with the cluster periphery. The radial existent of these boundary must be least than that of the cluster radius. Thus, the prescribed contours seems to be excellent tracer of the initial morphology of the stellar distribution.
\subsection{Cluster masses}
The mass of cluster ($M_C$) is computed through the value of the galactocentric distance $R_G$ as follows \citep{ki62},
$$M_C = \frac{4A(A-B)r^3_{t}}{G},$$
where $A$ and $B$ are Oort's constants valid for $R_{G}$, $r_{t}$ and $G$ are the tidal radius and gravitational constant \citep{st95}, respectively. These Oort’s constants of the open clusters subsystem could be easily expressed by their local values as \citep{pi07},
$$A_{0}=14.5{\pm}0.8~km~s^{-1}kpc^{-1},~~~B_{0}=-13.0{\pm}1.1~km~s^{-1}kpc^{-1}.$$
Furthermore, these local values are provided exact constant as follows,
$$A=A_{0}-A_{0} \delta{R_{G}}$$
$$A-B=A_{0}-B_{0}-2A_{0} \delta{R_{G}}$$
where $\delta{R_{G}}=(R_{G}-R_{G,0})/R_{G,0}$ and $R_{G,0}=8.5~kpc$.
The relative random error of the cluster mass as expressed as \citep{pi07},
$$\delta_{M_C} \approx 3 \delta_{r_{t}},$$
where $\delta_{r_{t}}= \epsilon_{r_{t}}/r_{t}$ and defined as the relative
random errors of the tidal radius.
\subsubsection{Half mass radius}
Half mass radius (HMR) of the clusters defined by a radial distance whose periphery contained the half mass of the cluster. The HMR of any cluster is entirely different from the the half physical radius of the clusters. The periphery of HMR is divided the whole cluster regions into the two equal mass parts such as (i) the central part and (ii) the halo part. Other hand, central part within the periphery of half physical is contained the one-third area of the remaining area. Thus, the cluster HMR seems to be more than its half physical radius. The prescribed HMR value is useful to estimation of the dynamical relaxation time of the studied cluster. 
\section{Dynamical properties of the cluster}
The spatial position of members is changed due to the stellar encounters among them and these encounters lead equi-partition of energies among the cluster members. The data incompleteness fraction is necessary to derive the dynamical properties of the cluster due to the fact that the it is not always possible to detect all the stars present in the CCD frame, particularly towards fainter end \citep{jo14}. These incompleteness test are performed through $ADDSTAR$ routine in $DAOPHOT$. In this routine, randomly selected artificial stars of known magnitudes and positions are added in the original frames. The cluster characteristics remains unchanged by adding few hundred stars, which probably $10-15 \%$ of the detected stars in the original frame. This said procedure are performed 5-10 times for each original image. However, it may also becomes a cause of error for the dynamical properties of the clusters. Thus, there are two contributions to the errors of dynamical properties: uncertainties on the correction factor sand uncertainties on the number of stars \citep{ha08}. Moreover, the correction factor effectively reduce the influence of data incompleteness. On this background, the dynamical behaviour of any cluster is understood as follow,
\subsection{Luminosity function}
The luminosity function (LF) is expressed as total number of cluster members in different magnitude bins \citep{jo14}. These members in various magnitude bins are computed through the following relation \citep{ni02},
$$N_{i}=\frac{N_{0}}{CFBC}-N_{F},$$
where $N_{i}$, $N_{0}$ and $N_{F}$ are the cluster members of the cluster, number of observed stars in the cluster and field regions respectively and $CFBC$ is the completeness factor for the considered MS brightness in the selected magnitude bins of any photometric band.
The K-band luminosity function (KLF) seems to be a powerful approach to understood the IMF of young embedded star clusters on the behalf of several studies \citep{oj04,sa07}. The KLF of these objects may be obtained through following relation \citep{pa08},
$$\frac{dN(K)}{dK} \propto 10^{\alpha K},$$
where $\alpha$ is the slope of power law, whereas $dN(K)/dK$ is total number of stars per unit $K-$magnitude bin. 
\subsection{Mass Function}
In the initial process of star formation within OCs, an amount of the stellar mass within in per unit volume known as the Initial mass function (IMF), which is very effective to determine the subsequent evolution of cluster \citep{kr02}. The direct measurement of IMF does not possible due to the dynamical evolution of stellar systems though we have been estimated the mass function (MF) of cluster. The MF is defined as the relative numbers of stars per unit mass and can be expressed by a power law $N(\log M) \propto M^{\Gamma}$. The slope, $\Gamma$, of the MF can be determined from
$$ \Gamma = \frac{d\log N(\log\it{m})}{d\log\it{m}} $$
where $N\log(m)$ is the number of stars per unit logarithmic mass. The masses of MPMs can be determined by comparing observed magnitudes with those predicted by a stellar evolutionary model if the age, reddening, distance and metallicity are known. The classical value of MF slope is -1.35 \citep{sa55}.
\subsection{Mass segregation}
This effect may be arisen due to the dynamical evolution or imprint of star formation or both \citep{ya08}. Through the process of Mass segregation, the stellar encounters of the cluster members are gradually increased. The stellar encounters among of members of clusters gradually lead to an increased degree of energy equipartition throughout the lifetime of the cluster \citep{ya13}. Furthermore, the higher-mass cluster members gradually sink towards the cluster center by transferring their kinetic energy to the more numerous lower-mass stellar members of the cluster \citet{ya13}. This phenomena is also statistically verified through Kolmogorov-Smirnov test.\\
To understand the mass segregation phenomena, the members of the clusters are subdivided into two/three groups of different stellar magnitude ranges.  We have plotted the two dimensional diagram of the normalized commutative frequency with the radial distance of the cluster. The said normalized frequency is defined as the ratio commutative stellar numbers to the total number of members within the given magnitude range. The convex type plot of these defined groups indicates the mass segregation phenomena within the studied cluster. 
\subsection{Dynamical relaxation time}
The relaxation time ($T_{e}$) of a cluster is defined as the time in which the cluster needs from the very beginning to build itself and reach the stability state against the contraction and destruction forces, e.g. gas pressure, turbulence, rotation, and the magnetic field \citep{ta05}. IN other words, this time ($T_e$) is also defined as the time in which the individual stars exchange energies and their velocity distribution approaches a Maxwellian equilibrium and mathematically is expressed as \citep{sp71},
$$T_e= 8.9 {\times} \frac{R^2_h \sqrt{N}}{log(0.4N)\sqrt{<m>}}$$
where $N$ is the number of cluster members, $R_h$ is the radius containing half of the cluster mass and $<m>$ is the average mass of the cluster stars. Both theory and simulations show that significant mass segregation among heaviest stars in the cluster core occurs in the local relaxation time, but affecting a large fraction of the mass of the cluster requires a time comparable to the average relaxation time averaged over the inner half of the mass \citep{in85,ch90,me97,ni02}. If, said value is found to be very low compare to the cluster age, then the studied cluster is said to be  dynamically evolved system. Similarly, the cluster is defined as the dynamically relax for other cases.
\section{The membership procedure for the cluster system}
Whenever, we have to looked towards a small portion of the Sky through scientific-eye i.e. telescope, then we are obtained many stellar enhanced region in the somewhere of Sky. These stellar enhanced region of the Milky-way must be possible stellar clusters. These clusters are mostly influenced by the foreground and background stars. The nature of field star sequences can only be clarified with a difficult membership analysis procedure \citep{vi04,mon10}. As a result, it is needed to define a factor/procedure for describing the membership reasons/status for detected stars within the precised clusters. Since, we have knowledge about position, mean proper motion and observed stellar magnitudes within various photometric bands, therefore, these information may be indicated the possibility of a star belonging to a particular cluster region. This said possibility must be defined by the membership probabilities of star, which are calculated by the various statistical approaches based on the known stellar parameters. In the simple words, the probability associated with the position is identified as the spatial probability, whereas that related to the proper motion is referred as the kinematic probability. Similarly, there are some statistical cleaning procedure are based on the observed colour-excess and stellar magnitudes and defined as the statistical probability. The criteria based on these probabilities is expected to yield the most probable members of a cluster. The members of open clusters have the same age, distance, proper motion and spatial velocity, hence they follow a pattern in the colour-magnitude diagrams and proper motion plane. The location of these most probable members in cluster’s CMD’s then becomes a tracing tool of the best visual fitted isochrones \citep{jo15a}. As such, the probabilities needed for the estimation of membership of stars within a clusters may be described as follows.
\subsection{Spatial probabilities}
These probabilities are statistical values which are computed by using the values of stellar radial distance. The basic assumption of this type probability that the probability of stars at the center of cluster is 1, whereas 0 for the stars of cluster periphery. Furthermore, said probability is considered to be 0 for the outsider stars from the cluster periphery. These probabilities are also subdivided into various categories according to the assumed variation nature of resultant probability as follow,
\subsubsection{Linear spatial probability}
The linear decrements of stellar probabilities with the radial distance is the basic principle of this type probability. For this, the probability of the detected stars has decreased in the relative fraction of radial distance to the cluster radius and it is written in the form of following mathematical expression \citep{jo14},
$$p_{s_{l}}=1-\frac{r_i}{r_{cluster}},$$
where $r_i$ is the radial distance of $i^{th}$ star from the center. Thus, resultant probabilistic values are dependent on the estimated center of the cluster, whereas independent from their dynamical parameters. The uniform decrements of the probabilistic values with the radial distance is unique property of its. Similarly, the values are unaffected by the cluster size.
\subsubsection{Non-linear spatial probability}
In this probabilistic approach, said values have decreased in the term of square of relative fraction $r_{i}/r_{cluster}$. As a result, the resultant formula have been obtained as,
$$p_{s_{non}}=1-\left(\frac{r_i}{r_{cluster}}\right)^2.$$
This term is representing the variance of the probability according to the square law.
\subsection{Photometric probability}
The photometric probability ($P_{ph}$) is computed with reference to the blue and red limits in the $B-V$ vs $V$ CMD. The blue sequence may be defined by using empirical zero-age-main-sequence (ZAMS) colours \citet{sc82} shifted in magnitude and colour to visually match the cluster sequence. The red sequence is defined by a shift of -0.75 mag in $V$ and a shift of $0.042$ mag in $B-V$ in order to account for unresolved MS binaries \citep{ma74, kh04}. Stars with B-V colours lying within the binary sequence of the MS are assigned $P_{ph}$ of 1 and are probable members of the cluster, while stars deviating along either direction are assigned a probability as,
$$p_{ph} = exp \{ -0.5 {\times} \left[ \Delta (B-V)/{\sigma_{B-V}} \right]^2 \}$$
where $\Delta (B-V)$ is the difference between colours from blue or red colour limits and $\sigma_{(B-V)}$ is the photometric error in colour.
\subsection{Kinematic probability}
It is well known that the proper-motion (PM) of member stars of a cluster usually differs the values from the mean-proper motion of the cluster. As a result, the  deviation of proper-motion of a member star in both $RA$ and $DEC$ directions with respect to the mean-proper motion of cluster provides a measure of its kinematic probability. The said proper motion values are provided different probabilistic values for a star on the behalf of adopted statistical approaches. We may be seen several membership approach in the literature. Here, we are discussed some well known approach following as,
\subsubsection{Iterative kinematic probability}
In this procedure, the mean-proper motion values of the detected stars (which are laying within cluster periphery) has computed by 3$\sigma$ clipping with iterative procedure. The mean and $\sigma$ values of such proper motion of each cluster have been determined in both the directions and are used to separate the stars not lying within the 3$\sigma$ value in both RA and DEC directions. Using iteration procedure, the stars lying outside 3$\sigma$ value from the mean value of proper-motion of cluster/stellar-system in both directions have been completely eliminated and the remaining stars are used to find the mean proper motion of the clusters. The kinematic probability of stars has been derived using these  prescribed mean proper motion values through the relation given as \citep{kh04}, 
\begin{equation}
p_k=exp{\left[-0.25\{(\mu_{x}-\bar{\mu_{x}})^2/\sigma_{x}^2 + (\mu_{y}-\bar{\mu_{y}})^2/\sigma_{y}^2\}\right]}
\end{equation} 
where $\sigma_{x}^2=\sigma_{\mu x}^2+\sigma_{\bar{\mu x}}^2$ and $\sigma_{y}^2=\sigma_{\mu y}^2+\sigma_{\bar{\mu y}}^2$. The $\mu_{x}$ and $\mu_{y}$ are the proper motion of any star in the direction of $RA$ and $DEC$ respectively, while $\sigma_{\mu x}$ and $\sigma_{\mu y}$ are the uncertainties in measurement of the proper motion of stars along the respective directions.
\subsubsection{frequency related probability}
In the past, the kinematic probabilistic algorithm through frequency distribution had been developed for members having proper motions of different observed precisions \citep{st80,zh90}. In later work, the correlation coefficient of the field star distribution to the set of parameters describing their distribution on the sky \citep{zh94}. The construction of frequency distributions have proposed for the cluster stars ($\phi^{\nu}_c$) and field stars ($\phi^{\nu}_f$) \citep{ba98}. The frequency function for the $i^{th}$ star of a cluster can be written as follows,\\
$\phi^{\nu}_c=\frac{1}{2\pi\sqrt{(\sigma^2_c+\epsilon^2_{xi})(\sigma^2_c+\epsilon^2_{yi})}}~{\times}~ exp~ \left\{ -\frac{1}{2} \left[ \frac{(\mu_{xi}-\mu_{xc})^2}{\sigma^2_{c}+\epsilon^2_{xi}} + \frac{(\mu_{yi}-\mu_{yc})^2}{\sigma^2_{c}+\epsilon^2_{yi}} \right] \right\},$\\
where $\mu_{xi}$ and $\mu_{yi}$are the PMs of the $i_{th}$  star, while $\mu_{xc}$ and $\mu_{yc}$ are the cluster’s PM centre. $\sigma_{c}$ is the intrinsic PM dispersion of cluster member stars, and ($\epsilon_{xi},\epsilon_{yi}$) are the observed errors in the PM components of the $i_{th}$ star. The frequency distribution for the $i_{th}$ field star is as follows,\\
$\phi^{\nu}_f=\frac{1}{2\pi{\sqrt{{(1-\gamma^2)}(\sigma^2_{xf} +\epsilon^2_{xi})(\sigma^2_{yf}+\epsilon^2_{yi})}}}$\\ $~~~~~~~~~~exp\left\{-\frac{1}{2(1-\gamma^2)} \left[\frac{(\mu_{xi}-\mu_{xf})^2}{\sigma^2_{xf}+\epsilon^2_{xi}} - \frac{2\gamma(\mu_{xi}-\mu_{xf})(\mu_{yi}-\mu_{yf})}{\sqrt{(\sigma^2_{xf} +\epsilon^2_{xi})(\sigma^2_{yf}+\epsilon^2_{yi})}} + \frac{(\mu_{yi} -\mu_{yf})^2}{\sigma^2_{yf}+\epsilon^2_{yi}} \right] \right\},$\\
where $\mu_{xf}$ and $\mu_{yf}$ are the field PM centre, while $\epsilon_{xi}$ and $\epsilon_{yi}$ the observed errors in the PM components. In addition, $\sigma_{xf}$ and $\sigma_{yf}$ are the field intrinsic PM dispersion values and corelation coefficient $(\gamma)$ is calculated by,
$$\gamma=\frac{(\mu_{xi}-\mu_{xf})(\mu_{yi}-\mu_{yf})}{\sigma_{xf} \sigma_{yf}}.$$
The spatial distribution may be excluded due to the small observed field of stars \citep{ya13}. The distribution of detected stars within observed field can be calculated as,
$$\phi=(n_c \phi^{\nu}_c)+(n_f \phi^{\nu}_f)$$
where $n_c$ and $n_{f}$ are the normalized numbers of stars for the cluster
and field ($n_c+n_f=1$). Finally, we have defined the membership probability for the $i^{th}$ star following as,
$$P_{\mu}(i)=\phi_c(i)/\phi(i).$$
This procedure is believed to be good indicator of the cluster and filed star separation. Since, the contamination by background and foreground objects can not be avoidable during observations, therefore, the effectiveness of the membership determination would be estimated by following relation \citep{sh96},
$$E=1-\frac{N \sum_{i=1}\limits^{N} {[ P_{i}(1-P_{i}) ]} }{{\sum_{i=1}\limits^{N}}P_{i}{\sum_{i=1}\limits^{N}}(1-P_{i})},$$
where $N$ is the total number of stars under consideration for the determination of the membership probability and $P_{i}$ indicates the probability of $i^{th}$ star of the cluster. In addition, the resultant larger value of $E$ represented larger effectiveness in the determination of membership probability of the stars.
\subsection{Statistical probability}
This type probability are generally based on the comparison of the cluster and field CMDs. According to this method, we removed all cluster stars in a colour-magnitude plane, which fall within a assumed size grid cell. In addition, the field region is taken in such a way that the area of field region must be equal to the area of the cluster region. These equal areas are needed due to symmetric distribution of field stars within observed region of the Sky. The field region is also chosen in two ways: (i) the loop region around the cluster periphery and (ii) one or two degree off from the coordinate of cluster centre. The remaining stars through this approach are called as the statistically cleaned members of the cluster. The statistical probability of above said members is considered to be 1, whereas 0 for other members. The various statistical probabilistic approach are described as below,
\subsubsection{Fixed-grid procedure}
The decontaminated CMDs have been used to investigated the nature of star cluster candidates and derive their astrophysical properties through various developed astrophysical tools \citep{bo07,bi11}. It is also believed that the 2MASS photometric CMD through field star decontamination has shown to constrain age and distance significantly more than the observed photometry, especially for low-latitude and/or bulge-projected clusters \citep{bi08,bi11}. In this procedure, the detected stars within the cluster periphery are divided into multi-dimensional (magnitude and colour axis) grid of cells. One dimension is referred as the magnitude, whereas others are colour. In the next step, total stellar (members+filed) number $n_{total}$ within cluster periphery is computed for each cell and also estimated the field stellar number $n_{fs}$ in corresponding cells of field region. Finally, the similar $n_{fs}$ stars has been subtracted from the cell of cluster region and these remaining stars may be used for further analysis. 
\subsubsection{field star dependent grid procedure}
In this procedure, every field star has its own grid with fixed dimension in such a way that the coordinate of field star becomes centre of its grid. The one dimension is magnitude and others are colours. Furthermore, the dimensions of these grids are also intersected to each other. In addition, we have identified those cluster stars, which are laying within assumed dimensions of the grid of field star. After it, the colour-magnitude distances of these stars have been computed from the field star and stars can be arranged in the increasing order of said distances. The star, having shortest distance from the field star, is eliminated from the cluster membership. This described procedure is iterated for every field star. Thus, one star from the cluster region has eliminated for a field star. The remaining stars are defined to be statistically cleaned  members of the cluster.
\section{The kind of cluster Members through photometry}
The stars are distributed overall the CMD and they may be further divided into various sub-types according to their position in the plane of CMD. The brief description about these types as follow,
\subsection{Early type members}
These are O to late-B type stars which can be satisfied the following conditions \citep{li14},\\
(i) $V \preceq 15$, $0.2 \preceq (B-V)\preceq 0.7$, $-1.0 \preceq (U-B) \preceq 0.5$, $E(B-V) \succeq 0.5$ and $-1.0 \preceq$ Johnson’s $Q  \preceq -0.1$;\\
(ii) an individual distance modulus between $(M_O-M_V)_{cl}-0.75-2.5\sigma_{M_O-M_V}$ and $(M_O-M_V)_{cl}+2.5\sigma_{M_O-M_V}$ to take into account the effect of binary members and photometric errors \citep{su99a,ko10}, where $(M_O-M_V)$ and $\sigma_{M_O-M_V}$ are the mean distance modulus and the width of the Gaussian fit of the distance modulus, respectively. The  ionization front (IF) is generated due to the ionization of cluster surrounding by UV radiation of these stars \citep{pa08}. The IF drives a shock into the pre-existing molecular clumps and compresses it, consequently the clumps become gravitationally unstable and collapse to form next-generation stars \citep{el98}. In the case of anomalous reddening within the cluster, the intrinsic magnitudes of stars, having spectral type earlier than $A_0$ were obtained as
follows \citep{pa08},
$$V_0=V-[3.1 {\times}E(B-V)_{min} + R_{cluster} {\times} \Delta{E(B-V)}],$$
where $\Delta{E(B-V)}=E(B-V)_{*}-E(B_V)_{min}$ and $E(B-V)_{*}$ is $E(B-V)$ value of individual stars estimated from the $Q$ method. These stars are used for estimating the reddening value of the cluster.
\subsubsection{Be stars}
The early type stars have been loosing their masses through the phenomena of MS evolution and this fact is found to be maximum in those young open clusters, which have age to be $10-25~Myrs$ \citep{fa00}. It is believed that about $20{\%}$ of the early type stars $<B5$ of the young open clusters are showing Be-phenomena \citep{ma99}. However, this phenomena may be overabundance ($>30 \%$) due to presence of Herbig Be stars, which also associated with nebulosity and show large near-IR excess \citep{su06}. The NIR excess stars with H$\alpha$ emission stars are important targets to study the effect of environment on the disc around the young stellar objects (YSOs). 
\subsection{$K_s$ excesses stars}
Before occupying the position in the ZAMS, stars surrounded by optically thick material consisting of an infalling envelope and accretion disc \citep{bo06a}. Disc strength as measured by excesses in intrinsic colours such as $(H-K)$ or $K_s$. These colours are depend on the stellar mass and radius, and disc properties such as accretion rate and geometry \citep{al99}. The intrinsic $K_s$-excess stars are those which having the foreground-reddening corrected colour $(H-K_s)$ redder than the $OV$ reddening vector \citep{bo06a}. It is emphasized that the $K_s$-excess stars  are brighter than $J=13~mag$. The $K_{s}$-excess stars of the 14 Myr open cluster NGC 4755 do not occupy the classical disc locus of unreddened T-Tauri stars \citep{bo06a}, however theoretical models predicts classical full discs \citep{la92}.
\subsection{UV excess and acceleration rate of PMS stars}
The disc evolution process around PMS stars may be understood through the mass acceleration rate \citep{li14}. The large amount of UV excess may be obtained due to intrinsic properties of late-type PMS stars rather than to inappropriate reddening corrections \citep{li14}. The limit of $UV$ excess from chromospheric activity is seems to be about $-0.5$mag \citep{re00}. The accretion luminosity $L_{acc}$ of these stars may be estimated the through the relation \citep{gu98},
$$log(L_{acc}/L_{\odot}) = 1.09 log(L_{U, exc}/L_{\odot})+0.98,$$
where $L_{U,exc}=L_{U,0}-L_{U,exp}$. By using a bandwidth (700$\AA$)and zero magnitude flux of $4.22 {\times} 10^{-9}~erg~cm^{-2}~{\AA}^{-1}$ for the Bessell $U$ band \citep{co00}, the luminosity ($L{U,exp}$ and $L_{U,0}$) have been computing from expected magnitude of the photospheric colour of MS stars ($U_{exp}$) and the extinction corrected magnitude of stars with $UV$ emission ($U_{0}$). The values of radius ($R_{PMS}$) and mass ($M_{PMS}$) and the accretion luminosity are used to estimate the mass accretion rate ($\dot{M}$) through following relation \citep{gu98,ha98},
$$\dot{M}=L_{acc}R_{PMS}/0.8GM_{PMS},$$
where $G$ is the Gravitational constant.
\subsection{Blue straggler stars}
Blue stragglers are those stars which are seem to stay on the main sequence longer than expected time through the standard theory of stellar evolution \citep{ah07}. Furthermore, these high mass objects seem to belong to the thick disc or halo population on the basis of their metallicity or kinematics \citep{ah07}. These said members can be identified through the relation between the $(B-V)$ index of the turnoff and the metallicity of the coeval, metal-poor, thick disc and halo populations \citep{ca94}. These objects appear to be above the turnoff leads as a horizontal-branch stars of post-main-sequence stars \citep{ah07}. Furthermore, these members are also located above the turnoff, blue-ward direction of the CMD and on or near the ZAMS \citep{ah07}. The analysis of four lithium-deficient stars indicates that stragglers produced by mass exchange in close binaries \citep{ry01}. Spectroscopic analysis of radial-velocity studies of metal-poor field blue stragglers suggest that the observed abundances (including the lack of lithium) are consistent with stable mass transfer during the asymptotic-branch stage of the primaries \citep{car05}.
\subsection{Giant members}
The Giant members are mostly found in the intermediate and old age open star clusters. These members are located in the redder side of the main sequence in the CMD. These stars are those stars which are evaporated from the main sequence due to the dynamical evolution. The possible cause of their evaporation from the MS may be fast evolution rate due to their heavy mass. In addition, the situated above and right side of the turnoff point of best fitted isochrone. Moreover, Giant members are the brighter members of the cluster. The position of Giant members with the MS stars gives the opportunity of better visual fitting of the theoretical isochrone of known metallicity.
\section{Discussion about the recent studies}
The wast work is carried out in the field of star cluster properties. The known cluster parameters are assembled in two comprehensive and frequently updated databases, WEBDA\thanks{www.univie.ac.at/webda} and the``Catalog of Optically Visible Open Clusters and Candidates" (DAML02)\thanks{www.astro.iag.usp.br/~wilton} \citep{bi11}. These studies are carried out on the basis of various statistical procedures as discussed in the earlier sections of this manuscript. The parameters of cluster $NGC~6866$ is also estimated through Basian analysis \citep{ja13}. The general discussion about cluster properties of some clusters as given below,
\subsection{Spatial properties}
The spatial properties of the clusters have been obtained from various photometric catalogues, such as 2MASS \citep{sk06}, PPMXL \citep{ro10} and other available $UBVRI$ database. Basically, the spatial properties of the clusters are varies with the the nature of gathered data. The Integrated $BVJHK_{s}$ parameters and the luminosity function of 650 galactic open cluster represented by Kharchenko et al 2009 \citep{kh09}. The fitting of King's model to open clusters creates difficulties due to the relatively poor stellar population (compared to globular clusters) and from the higher degree of contamination by field stars in the Galactic disk \citep{pi07}. Though, the stellar crowding effect may be helped to identify the clusters but some time false estimation may be occurred. Recently, Dolidze 39, BH 79, and Ruprecht 103 seems to be field fluctuation instead of the clusters \citep{bi11}. The analysis of concentration parameter of the clusters indicates that the angular size of the coronal region is about 6 times the core radius \citep{nil02}. The limiting radius of the clusters may vary for individual clusters between about 2$R_{core}$ and 7$R_{core}$ \citep{ma07}. The cluster radius is found to be larger in infrared bands compare to the optical bands \citep{sh06}.
\subsection{Importance of Membership analysis}
The cluster mass could simply be derived from the summation of masses of individual members \citep{pi07}. The identification procedure of cluster members is highly influenced by the limiting magnitude of detected stars and uncertainty in the determination of these magnitude. Thus, sufficiently accurate and uniform mass estimates for a significant number of OCs in the Galaxy will obtained through the deeper surveys and an increasing accuracy of kinematic and photometric data \citep{pi07}. Since, the broadness of stellar scattering has been reducing through the field star decontamination, therefore we gets a opportunity to obtained more accurate results. For example, the MF slopes $x=1.65{\pm}0.20$ and $1.31{\pm}0.50$ are derived for NGC 637 and 957 by considering the corrections of field
star contamination and data incompleteness \citep{ya08}. 
\section{Unresolved mysteries of the recent studies}
The identification of exact members of any cluster is most crucial and mystical work due to the least information of them. There are several approaches to their identification but each approach has its own limitations. It is believed that the most probable members of cluster may be satisfied the pattern of theoretical isochrones but their are very limiting work for this purpose. The data incompleteness is a major factor to influence the cluster characteristics. This said factor is highly changed the MF values for the fainter members of the clusters leads the deviation of slope of MF from the classical Salpeter value. This variation may leading the revolutionary changes within the cluster morphology. However, the ADDSTAR routine has been utilized to overcome the effect of said incompleteness but it is highly dependent on the detected stars through the applied photometric techniques. Presently, there are no alternative approach to verify the precision of applied completeness factors.\\
Mostly, the dynamical studies are carried out either in visual $V$-band or in 2MASS-band, but the cross verification of their results is further needed due to not matching results from these studies. The stellar enhancement provides the cluster existence which is highly contaminated by the presence of field stars. In addition, the detection of stars is highly effected by the stellar crowding within the cluster region, which leads low detection rate of field stars of the cluster compare than surrounding. It is also noticed in several work that the cluster periphery has been computed before applying the completeness correction. Since, the completeness correction of field and cluster region are found to be different, therefore, it is needed to develop a procedure of estimating the cluster radius, which may be linked with the completeness corrections. The estimation error of stellar magnitude creates major difficultly to determine the characteristics of fainter members of the cluster. The deep observations may reduces this difficulty but there are needed a methodology for increasing the accuracy of these members.\\
The turn point of MS play a major role in the estimation of distance, age of the cluster through the theoretical isochrones. The identification of this point is highly dependent on the stellar distribution of CMD and appeared width of the MS in said CMD plane. Since, the said turn point of the cluster is mostly defined by the visual inspection of the obtained stellar distribution in the CMD, therefore, it is needed to constrain a procedure/model for determining the said turn point, which would be very effectively used to verify the turn point through the visual inspection. The synthetic CMD method and Bassian analysis had utilized to constrain the evolutionary isochrone through the observed data points. For these purpose, pre-determined cluster members are used, therefore, the effect of field stars and spread stellar distribution is also demanded some attention.\\
The determined distance and age of the cluster is also related with the metallicity. Generally, the value of metallicity has been selected through the pre-consideration. There are no photometric technique for determining this. The spectroscopic observations of massive stars of the clusters are utilized to determine the metallicity of the cluster. Such type spectroscopic observations of the clusters are very limited and also the the spectroscopic study of the very fainter members are not available. On this background, the accurate measurement of the metallicity of the cluster is not available at the present. Since, the known cluster parameters depend on the pre-assumed metallicity of the cluster, therefore, the derived fundamental parameters of the cluster may be far away from the reality. On this background, the deep spectroscopic observations would be needed to determine the exact metallicity of the clusters.\\
It is well known fact that the cluster radius is determined by considering the the spherical shape of the cluster, which is far away from the reality. It is needed to use an equation of other possible shape to determine the cluster extent, which may be change the present understanding towards the clusters morphology. Since, the cluster are placed within the three dimensional (3D) space and we are determined their extension by considering them as a two dimensional objects, therefore, the new 3D models are needed to constrain the clear picture of cluster extent. The limiting radii and core radii of the clusters are linked with the density profile, therefore, the identification of exact cluster center becomes most important due to fact that it may be occurred different for the non-spherical clusters. Furthermore, the isodensity contour may become signature of the identification of the said center. The fraction of massive stars of a cluster may also signature of its ongoing evolutionary process leads dynamical behaviour and stage of the studied cluster.\\
The mean proper motion of the cluster is another site to understand the dynamical picture. The analytic study may open an opportunity to understand the future of cluster and varying characteristic of the clusters with time. Since, the known proper motion values of the clusters are least reliable due to larger uncertainty in their estimation, therefore, the comprehensive ground based observations needed for this purpose. Such type study may further leading to analysis the gravitational field within the cluster. The varying mean proper motion of various stellar mass bins may be useful to verify the mass segregation phenomena within the cluster. It is also needed to study about the possible changes of the proper motion of the cluster members with time. Its deep analysis may also helpful to estimate the variation of evaporation rate of cluster members.\\
The interstellar dust grains/clouds can be identified through the the analysis of TCDs. Since, said TCDs are highly depends on the detected stars in the various bands. The observed magnitudes are varied due to the the quantity of the dust in the direction of cluster, which may reduce the brightness of the stars. Since, the interstellar dust/cloud has uniformly distributed within the cluster region, therefore, the reduced fraction of the members are varied according to their mass. Thus, it is necessary to applying magnitude corrections on the detected stars. Such correction must be changed their distribution on the CMD plane leads changes in the fundamental parameters of the cluster. In addition, these TCDs are also effectively used to determine the stellar subcategory and their analysis may helpful to understand the star formation mystery within the cluster.\\
Besides these, there are several OSCs, for which, we have been very poor knowledge. The studies of this poor studied cluster and unknown cluster would be increased the database, which would be effectively utilized to constrain/verify the Galactic evolution models. The combined results of the clusters may provides information about their evolution process and its dependency on their size. The locations of OSCs can be effectively characterized the nature of density waves within the spiral arms. In conclusions, the wide database of the clusters seems to be the building blocks of proposed evolutionary model of the our Galaxy and the star formation sites. 
\section*{Acknowledgments}
This research has made use of the VizieR catalogue access tool, CDS, Strasbourg, France. The original description of the VizieR service was published in A{\&}AS 143, 23. GCJ is also thankful to AP Cyber Zone (Nanakmatta) for providing computer facilities. GCJ is also thankful to Shree Nilamber Joshi for providing the friendly environment in the my present working place, which becomes a milestone of my research work.
\bibliographystyle{model2-names}

\end{document}